RESEARCH

# inPHAP: Interactive visualization of genotype and phased haplotype data

Günter Jäger*, Alexander Peltzer and Kay Nieselt

## Abstract

**Background:** To understand individual genomes it is necessary to look at the variations that lead to changes in phenotype and possibly to disease. However, genotype information alone is often not sufficient and additional knowledge regarding the phase of the variation is needed to make correct interpretations. Interactive visualizations, that allow the user to explore the data in various ways, can be of great assistance in the process of making well informed decisions. But, currently there is a lack for visualizations that are able to deal with phased haplotype data.

**Results:** We present inPHAP, an interactive visualization tool for genotype and phased haplotype data. inPHAP features a variety of interaction possibilities such as zooming, sorting, filtering and aggregation of rows in order to explore patterns hidden in large genetic data sets. As a proof of concept, we apply inPHAP to the phased haplotype data set of Phase 1 of the 1000 Genomes Project. Thereby, inPHAP's ability to show genetic variations on the population as well as on the individuals level is demonstrated for several disease related loci.

**Conclusions:** As of today, inPHAP is the only visual analytical tool that allows the user to explore unphased and phased haplotype data interactively. Due to its highly scalable design, inPHAP can be applied to large datasets with up to 100 GB of data, enabling users to visualize even large scale input data. inPHAP closes the gap between common visualization tools for unphased genotype data and introduces several new features, such as the visualization of phased data.
inPHAP is available for download at http://bit.ly/1iJgKmX.

## Background

Combinations of genetic variants occurring on the same DNA molecule are known as haplotypes. The term haplotype was first used in 1967 in conjunction with the Human Leukocyte Antigen (HLA) system, a set of genes located close together on chromosome 6. This system of genes is important for determining tissue compatibility for transplants [1]. When studying haplotypes one distinguishes phased haplotypes and unphased genotypes. For a phased haplotype both the maternal and paternal alleles are known, either by directly inferring the information or using haplotype phasing tools. In contrast to that, for unphased genotypes the chromosomal origin for each allele is unknown.

Especially collecting and comparing single nucleotide variations (SNV) between different human populations has become of central interest. Abecasis *et al.* showed that human individuals have around $4 \times 10^6$ variants on average [2]. These variants can have great influence on genes, leading to malfunction or even complete loss of function and consequently to genetically related diseases such as cancer. To fully understand the mechanisms leading to disease, a catalog of all existing variants, especially of rare ones that are only seen in a single or very few individuals is required [2]. In addition, humans are diploid organisms, which means that they have two copies of each chromosome. Genes or other non-coding sequences constituted by two homologous chromosomes can be genetically very different.

Often the term haplotype is also used to refer to clusters of inherited single nucleotide polymorphisms (SNPs). By examining haplotypes, researchers wish to identify patterns of genetic variation that are associated with descent, phenotype or disease state. However, studying diploid, omni- or even polyploid organisms requires additional phase information, linking a specific genetic variation to its respective chromosome. Only by including such information one is able to understand the impact of genetic variations.

Furthermore, a widely used strategy in this context is to compare samples from several populations and to identify genomic loci or regions with significant genetic differentiation between these populations.

*Correspondence: guenter.jaeger@uni-tuebingen.de
Integrative Transcriptomics, Center for Bioinformatics, University of Tübingen, 72076 Tübingen, DE
Full list of author information is available at the end of the article

Many studies that genotype individuals have already been and are currently performed. The International HapMap Project [3] for example is an international consortium of scientists who catalog the complete genetic variation in the human genome. As of today more than 26.3 million SNPs have been listed in HapMap. Another example is the Collaborative Oncological Gene-environment Study (COGS) which tries to understand the genetic susceptibility of different hormone-related cancers [4, 5, 6, 7, 8]. Most haplotypes do not span more than one gene, so studying local relationships of SNPs is the most common use case.

Genome-wide association studies (GWAS) have been used successfully for dissecting the genetic causes underlying certain traits and diseases. Work by the Wellcome Trust Case Control Consortium[1] has identified variations-associated phenotypes ranging from malaria [9] to myocardial infarction (Myocardial Infarction Genetics Consortium, 2009) [10]. Typically, GWAS data are displayed using Manhattan plots, a type of scatter plot to display dense data, usually with non-zero amplitude. In GWAS Manhattan plots, genomic coordinates are displayed along the $x$-axis, and the $y$-axis represents the negative logarithm of the associated $p$-value for each polymorphism in the data set. Because strong associations have very small $p$-values, their negative logarithms will be the largest and visibly most prominent [11]. A number of tools or even whole suites are specifically designed to visually investigate variants, either separately or in their haplotype contexts. The SNP & Variation Suite [12] is a collection of analytical tools for managing, analyzing and visualizing genomic and phenotypic data. However, only well-established visualizations for SNP data are provided, most do not scale well with big data. Flapjack offers interactive visualization of large-scale genotype data with a focus on plant data [13]. Its emphasis is put on real-time rendering of the data and combining genotype data with phenotype or QTL data. Some genome browsers also offer additional visualization modes that allow visualization of genotype cohort data by agglomerating data from many individual genomes. Savant [14] in its latest version offers visualization for multi-individual genotype data sets by agglomerating SNPs from larger genomic regions and linking them with a linkage disequilibrium (LD) plot as originally introduced by Haploview [15].

While all described genotype and haplotype visualization tools so far mostly focus on showing raw data, Haploscope visualizes haplotype cluster frequencies that are estimated by statistical models for population haplotype variation [16]. Another example in this area is iXora [17], which is a framework for inferring haplotypes from genotyped population data and for associating observed phenotypes with the inferred haplotypes. It features statistical tests, such as Fisher's exact test, and visualization methods that help to study parental haplotype distributions or to spot unexpected distortions. These visualizations basically include line charts for haplotype frequency distributions as well as bar plots for haplotype visualization. The user can easily observe haplotypes, missing data, position of the markers on chromosome maps and colocalization with QTL.

In general, the analysis of haplotype data is a challenging scientific endeavor, since it involves the scalable processing of very large, heterogeneous, incomplete, and potentially conflicting data. Clearly, visualizing the data has been shown to aid in gaining better understanding of it. Furthermore, researchers wish to view all facets of haplotype data, including the spatial distribution of the loci along a chromosome, the specificity of the genotypes, the different frequencies of haplotypes in different subgroups, and possibly also correlation of occurring haplotypes. For this, static visualizations are insufficient, since such complex data needs to be addressed on many different levels, and here in particular interactivity is of utmost importance.

The challenges of visualizing haplotype data could be exacerbated when it comes to analyzing phased haplotype data that are for example derived from studies [18] such as the 1000 genomes project. Until today an interactive tool for the visualization of phased haplotype data has been missing. To fill the gap, we implemented inPHAP, short for (*in*teractive *P*hased *HAP*lotype Viewer). inPHAP can be used in several ways, ranging from the investigation of phased haplotypes or unphased genotypes on the single nucleotide level to the visualization of the data in a more general way showing the similarities and dissimilarities between several subject groups of interest. In the following, inPHAP and its features are presented, accompanied by a proof of concept application to data from Phase 1 of the 1000 Genomes Project.

## Methods

This section presents the general framework and the design choices we made for inPHAP.

inPHAP is an interactive visualization tool written in the JAVA programming language. It makes use of the general idea of iHAT [19], our previously published tool for the visualization and analysis of genome wide association (GWA) data. In iHAT we introduced the concept of interactive aggregation of subsets of the data in order to reveal hidden patterns that are not clearly visible when displaying the whole data set at once. Based on the concept of aggregating the information content of data based on meta-information, we implemented inPHAP, a new interactive visualization tool which is capable of visualizing unphased genotypes as well as phased haplotypes.

[1](http://www.wtccc.org.uk)

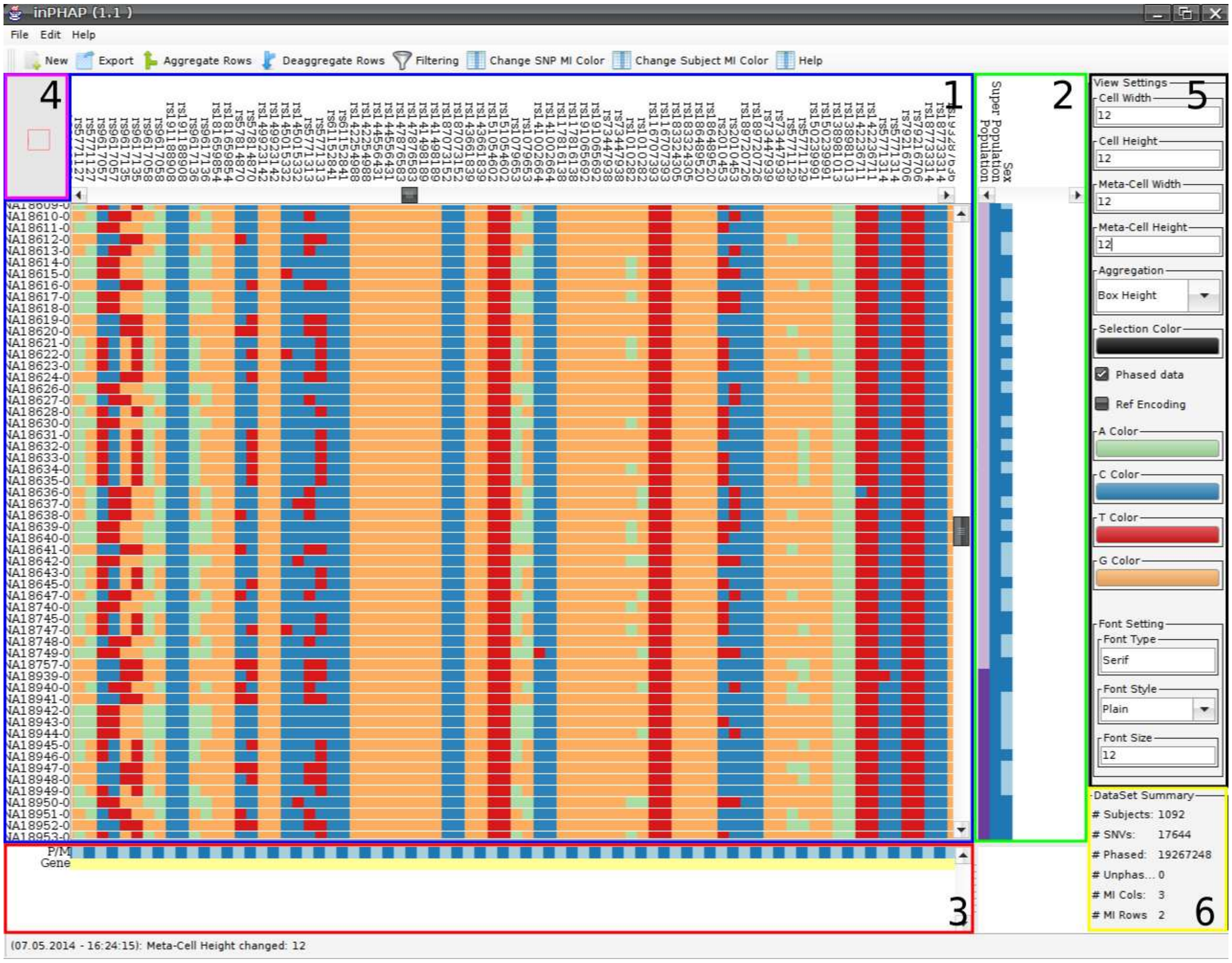

**Figure 1** The inPHAP graphical user interface. It consists of six components which are highlighted with boxes of different color. **Blue** (1): The haplotype visualization panel providing color-encoded base information for phased haplotype or unphased genotype data, **green** (2): the subject meta-information panel next to the haplotype visualization panel, **red**(3): the SNV meta-information panel below the haplotype visualization panel, **purple**(4): the overview panel, displaying the viewers current focus in the haplotype visualization panel, **black**(5): the settings panel, which allows the user to quickly change between settings, **yellow**(6): the data set summary panel, providing general information for the currently loaded data set.

In the following the design of the inPHAP tool itself, as well as its features are described in detail.

## The inPHAP Graphical User Interface

One of the key features of inPHAP is that it supports a broad range of interaction with the data. Therefore, we implemented a graphical user interface (GUI) which consists of six components (see Figure 1 for an overview of all the components): the haplotype visualization panel, the subject meta-information panel, the single nucleotide variation (SNV) meta-information panel, the overview panel, the settings panel, and last but not least the summary panel. The largest and most important component is the haplotype visualization panel located in the center of inPHAP. It consists of a heatmap-like haplotype visualization, together with row and column headers showing subject and the SNV identifiers, respectively. Detailed information on the visual representation of haplotype data is given in the *General Visual Encoding* section. The second component is the subject meta-information panel, which displays numerical and categorical meta-data of the subjects. Each meta-information type is represented as a single column in the subject meta-information panel and different color gradients for numerical data or maps for categorical data can be chosen by the user to distinguish sub-groups in the data. The SNV meta-information panel is used to enhance the haplotype visualization by displaying meta-information for variants. In the case of phased data for example, variants on the paternal and maternal chromosome can be distinguished. This information is then used to automatically create a meta-information row below the haplotype view with "P/M" as identifier to enhance identification of paternal and maternal alleles in the haplotype visualization panel. The fourth component in the upper left is the overview panel, an interactive zoomed out representation of the whole haplotype visualization. It shows the current view of the user in the haplotype visualization panel and gives an estimate of the proportion of the visualized data using a rectangle as visual clue. The settings panel on the right allows for quick changes of the most often needed settings. Here the user can change the way the data is presented. Amongst others, colors can be adjusted according to the users' needs and different visual representations for haplotype data are available. The last component is the data set summary panel. It provides general information for the current data set, including the number of subjects and SNVs in the data set as well as the number of different meta-information (MI) types, separated into "MI columns" and "MI rows" for subject and SNV meta-information. These panels are complemented by a button bar at the top of the GUI that provides convenient access to further useful and often needed functions, such as filtering, changing the subject or SNV MI color gradients or the export of the haplotype visualization. Additional functionality that is not available in the settings panel or the button bar is provided in the inPHAP menu bar. Furthermore, an information bar at the very bottom shows the last change made by the user. Thereby, it provides information on what has been changed and how this change affected the underlying data. A complete log of all interactions performed on the data is also available in the help menu located in the inPHAP menu bar.

## Data formats and structures

Data can be imported in inPHAP in two different formats: The VCF file format containing haplotype information for different subjects as separated columns and the IMPUTE2 format, the default haplotype text file format used by the IMPUTE2 program [18] to encode genotype information from the 1000 Genomes Project. The example files that have been used in our paper to demonstrate inPHAP have either been generated using SHAPEIT2 [20, 21] or BEAGLE [22, 23], which can both be used to infer phased haplotypes and are able to output the results in the IMPUTE2 or VCF file format. Since such files can get very large, implementation of the underlying data structures has been performed with respect to the overall memory consumption. In general, haplotype data consist of two different characters from the alphabet $\Sigma = \{A, T, C, G\}$, one character for the paternal allele and one for the maternal allele. In some cases also the character "$-$" is allowed, to indicate that no second allele is present. This is for example the case for many SNVs for the human X chromosome, especially for males. Encoding these characters as character primitives in Java would require 2 Bytes per character. For a dataset consisting of around $4 \times 10^6$ SNVs and about 1000 subjects this would lead to a memory consumption of $2 \times 4 \times 10^6 \times 10^3 \times 2 = 16$ GByte just for storing allele combinations. State of the art computers currently have between $8 - 16$ GBytes of RAM installed. To allow users to use inPHAP on their desktop computers, it was necessary to introduce a binary encoding of the haplotype data in order to reduce the amount of consumed memory. In inPHAP each character $c \in \Sigma$ is encoded using only two bits. With this strategy only 4 bits are needed to store the paternal and maternal allele for one SNV and subject. As a result inPHAP consumes for $4 \times 10^6$ SNVs and $10^3$ subjects only $(4 \times 10^6 \times 10^3)/2 = 2$ GByte for storing the raw allele combinations, which is 8 times less than using a naive memory storage approach.

To keep interactions smooth even on the lowest zoom level, where each cell of the haplotype visualization is $1 \times 1$ pixel in size, only those data that are needed for the currently visible submatrix are decompressed from their binary form. All other data is kept in the compressed form in memory. Furthermore, the visualization of the subject specific haplotypes has been optimized to perform very fast repainting. For this, each base $c \in \Sigma$ is rendered as a colored image in memory. When drawing the visible submatrix only already prerendered images are drawn, decreasing calculation and painting time to a minimum. To allow for smooth interaction with the visualization, selection boxes as well as different saturation values have also been implemented as pre-calculated images that can be drawn on top of the nucleotide images. With this strategy typical interactions, such as resorting the matrix, moving the sliders, or selecting specific columns or rows, do not require to recalculating the pre-rendered images but only repainting them in the current view. Changes that require a recalculation of the images, such as changing the color for the bases, then only requires to recalculate 4 images, which can be used multiple times for a single repaint event. Altogether, these mechanisms enable instantaneous updates of the haplotype visualization panel and smooth interaction in inPHAP.

In addition to haplotype data, meta-information data can be imported for subjects and for SNVs. Currently inPHAP accepts only tab-delimited text files with two header lines, with column names in the first header line and declaration of the type of data (categorical or numerical) for each column in the second header line, and subject and/or SNV identifiers in the first column.

On aggregated data, inPHAP utilizes a further visualization method to provide the user with feedback on the relative frequency of a certain nucleotide for the aggregated group of individuals in form of displaying a height of a bar within the respective cells. This can be changed by selecting the ''Saturation'' based visualization, which visualizes the most common SNV within the group by changing the color saturation from very low (= there are a lot of other SNVs within the group disagreeing with the shown SNV) to very high (= most of the SNVs within the aggregated group agree with the shown color), providing useful feedback as well for the user.

### General Visual Encoding

In the haplotype visualization panel there are two different visualizations available, one for phased data and one for unphased data. For phased data, each SNV is represented by two different columns, one for the paternal allele and one for the maternal allele. This design choice is motivated by the 1000 genomes data from Abecasis *et al.* who used two rows for each allele in their publication [24]. For unphased data only one column per SNV is needed. In addition, inPHAP offers two different color encodings for phased data and one for unphased data. In the default visual representation for phased data, each base is assigned a unique color. By default green is used for A, blue for C, red for T and yellow for G. Missing nucleotides, as it might be the case for males on the X chromosome are colored white. This encoding allows the user to compare different SNVs as well as to spot differences between the maternal and paternal allele quickly. The second visual representation for phased data is more convenient for visualizing differences to the reference base. If for one of the SNVs either the maternal or paternal allele differs from the respective reference base, then yellow color is used in the haplotype visualization panel, otherwise the respective cell is painted in blue. The third visual representation is more focused on unphased data, but can be applied to phased data as well. Here only one column is required for each SNV. If the phase is unknown, only three different cases can occur, namely homozygous and heterozygous SNVs as well as SNVs for which both alleles are equal to the reference base. Homozygous SNVs are colored red, while heterozygous SNVs are shown in yellow. If both alleles are equal to the reference the respective cell is colored green. For each of the three visual encodings, the default colors are selected based on ColorBrewer color maps [25], such that differences as well as similarities in the haplotype visualization panel can be spotted quickly. However, all colors can easily be changed in the settings panel to fulfill user specific needs. In case of a user defined selection of subjects of SNVs a colored border is drawn around cells in the haplotype visualization panel and the respective column or row identifiers are overlaid by a colored box. The default selection color is black, but it can also be changed by the user if needed.

In contrast to haplotype data, meta-information data is encoded in a different way. Here, for each meta-information the user can choose the appropriate color encoding. For numerical meta-data, the values are mapped directly to a color from the chosen color gradient. For categorical meta-data, first each category is assigned a unique numerical value. Then these numerical values are used for the selection of colors from the chosen color map.

### Interaction possibilities

*General interaction features*

inPHAP is a highly interactive tool, allowing the user to change the current view on the data in various ways. Interaction possibilities include the navigation along the subject (vertical) axis as well as along the SNV (horizontal) axis using the navigation bars. Furthermore, navigation is also possible using the overview panel. There, the current view is indicated by a red rectangle. This rectangle can be dragged to the desired location inducing a change in the position of the navigation bars in the haplotype visualization panel. Further interaction possibilities are zooming in two different dimensions, i.e. the width and height of each cell in the haplotype visualization panel can be adjusted. In addition, width and height of the meta-information cells can be changed separately from the visualization panel, allowing the user to see the meta-information assigned to subjects or SNVs even for very small cell sizes in the haplotype visualization. Width and height changes can be made either by using the settings panel or via the mouse wheel if the mouse is placed above the haplotype visualization panel or one of the meta-information panels, respectively. Subjects as well as SNVs of interest can be selected with the click of a mouse button on the respective identifier or via dragging over a series of identifiers. Selection thereby also affects the meta-information panels and the corresponding meta-information cells are highlighted as well. Furthermore, rows and columns in the haplotype visualization panel can be sorted according to the provided meta-information by double-clicking on one of the meta-information identifiers. For the sorting we use a stable sort. If the user for example chooses a meta-information group for sorting, the order of the elements that belong to the same subgroup in the chosen meta-information group is preserved. This allows users to sort according to different meta-information groups consecutively. These general interaction possibilities are assisted by several interactive filtering and aggregation methods, which will be explained in the following.

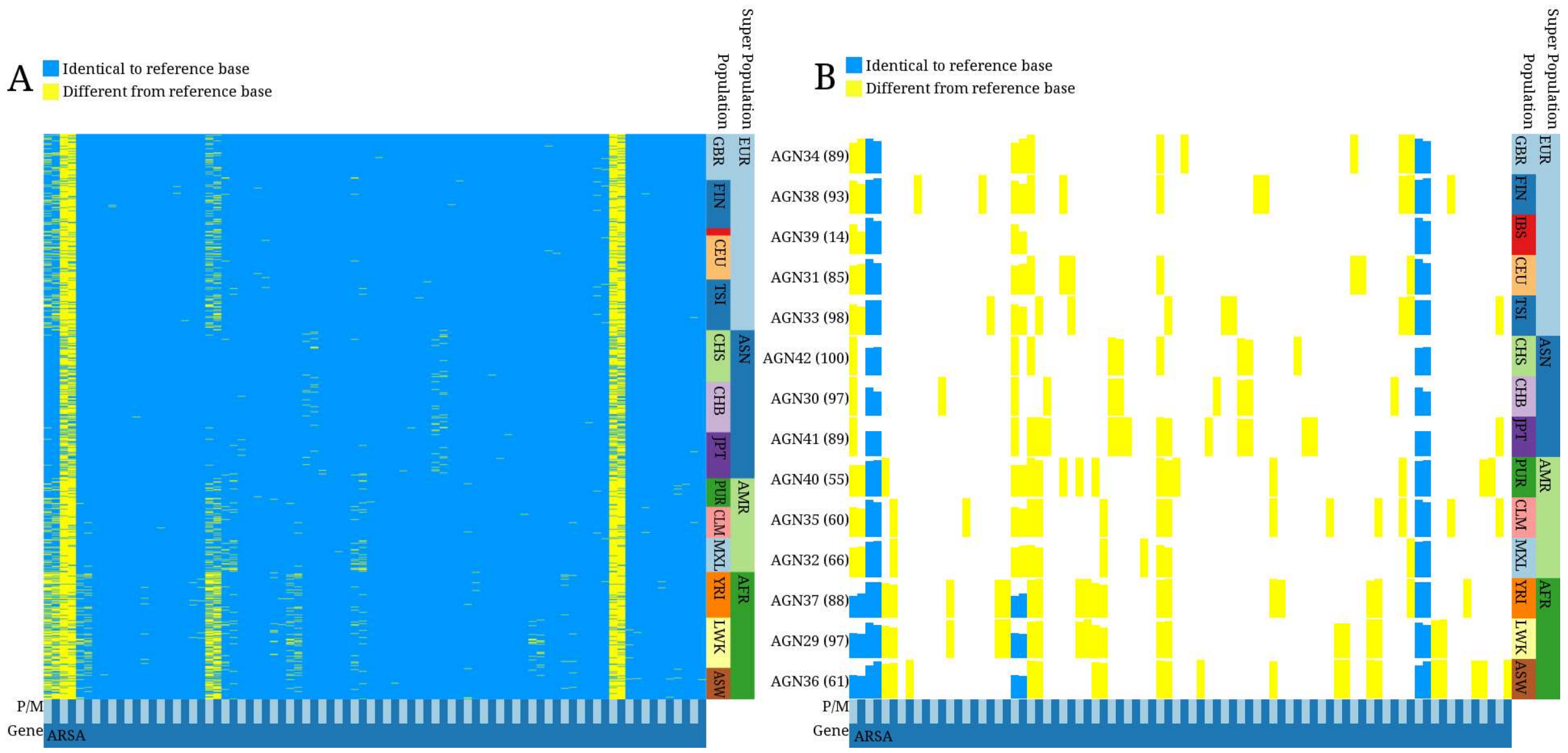

**Figure 2** Two inPHAP visualizations showing SNVs for the MLD associated gene *ARSA*. Data was taken from Phase 1 of the 1000 Genomes Project [24]. For both visualizations the reference color encoding was used. **A**: shows a non-aggregated view of the data, **B**: shows an aggregated view of the data. Individuals have been aggregated according to their population affiliation using the "minimum" aggregation method for SNVs. Bar heights represent rarity of the aggregated consensus base.
Population abbreviations: ASW, African ancestry in Southwest United States; CEU, Utah residents with ancestry from Northern and Western Europe; CHB, Han Chinese in Beijing, China; CHS, Han Chinese South, China; CLM, Colombians in Medellin, Colombia; FIN, Finnish in Finland; GBR, British from England and Scotland; IBS, Iberian populations in Spain; LWK, Luhya in Webuye, Kenya; JPT, Japanese in Tokyo, Japan; MXL, people with Mexican ancestry in Los Angeles, California; PUR, Puerto Ricans in Puerto Rico; TSI, Tuscani in Italy; YRI, Yoruba in Ibadan, Nigeria. Superpopulation abbreviations: AFR, African; AMR, Americas; ASN, East Asian; EUR, European.

*Filtering*
Filtering is a crucial step in the analysis of large data since it allows reducing the overall amount of data that has to be investigated by displaying only those variants that are of interest to the user. Consequently, data that is currently not of interest is removed from the view. If for example the user is interested in the variants that are shared by whole population groups rather than by only very few individuals, using a frequency filter can help in the selection of the respective SNVs and thereby reduce the overall amount of data that has to be visually assessed. To enable filtering in inPHAP, we implemented several different filter methods for single nucleotide variants. Filtering based on chromosomal location allows the user to concentrate on those SNVs that are located in a specific region on a chromosome, e.g. a gene or promoter region. If a list of interesting SNVs is already available, i.e. the user is interested in a specific haplotype, this list can be passed to inPHAP. Then only the intersection of SNVs in the given list with SNVs in the data set will be shown in the haplotype visualization panel. In addition, filtering based on SNV identifiers can also be done by providing a regular expression for the SNV identifier. We also included a frequency based filter, to show only those SNVs where the respective genotype frequency lies above or below a user-defined threshold. This is especially useful when the user wants to concentrate on rare variants only for example.

*Aggregation*
Using visualization to identify patterns in large data such as those from the 1000 Genomes Project is a challenging task, since structures often remain hidden when visualizing them on a global level. Therefore, methods to reduce the overall complexity of the data are needed to improve visual assessment of underlying patterns. In iHAT [19] we have demonstrated that aggregation is a rich technique when it comes to revealing hidden structures in the data. inPHAP allows the user to aggregate rows interactively, where for example meta-data can be used to guide this process. Especially for genotype as well as haplotype data where differences between whole populations or subgroups of populations are hard to compare, aggregation can help to unravel the hidden structures and thereby help to interpret the genetic differences. In inPHAP several different aggregation methods have been implemented, such as maximum, minimum or mean. A typical use case of aggregation of haplotype data would be to take subjects from a common group, e.g. from the same population, and look for differences in the haplotypes of these populations possibly revealing recombination events on a global level. In inPHAP the user can combine subjects of interest into subject groups by aggregating the corresponding haplotypes. These subject groups can either be based on user selection or on meta-information that has been additionally assigned to each subject. The aggregation of haplotypes is performed on a per SNV base. For each SNV the base with the highest frequency among the selected subjects is chosen as the consensus and the respective frequency is stored as an indication of how representative this base is given the underlying base distribution. In the haplotype visualization panel, aggregations can be encoded in two different ways, depending whether more attention shall be drawn to the consensus base itself or to the differences in SNV frequency in the combined subject group. If one is interested in the consensus base itself rather than in the differences in frequency between aggregated SNVs, aggregations can be represented as colored boxes where their saturation is adjusted based on the frequency of the consensus base. This visual representation is the default representation that was shown to work well on genotype data [19]. However, in a study conducted by Mackinlay it was shown that positioning along a common scale is more effective than saturation when comparing quantitative values [26]. inPHAP therefore offers an alternative way to represent aggregations. Instead of filled boxes, bars are drawn, whose color represents the consensus base and the height of the bar displays the underlying consensus base frequency. With this second visual encoding, differences in frequency stand out more clearly, which is especially useful for the comparison of maternal and paternal allele frequencies. Aggregated individuals are assigned a new identifier in the haplotype visualization panel constructed from the prefix “AGN” followed by a number. This number corresponds to the number of individuals included in the aggregation.

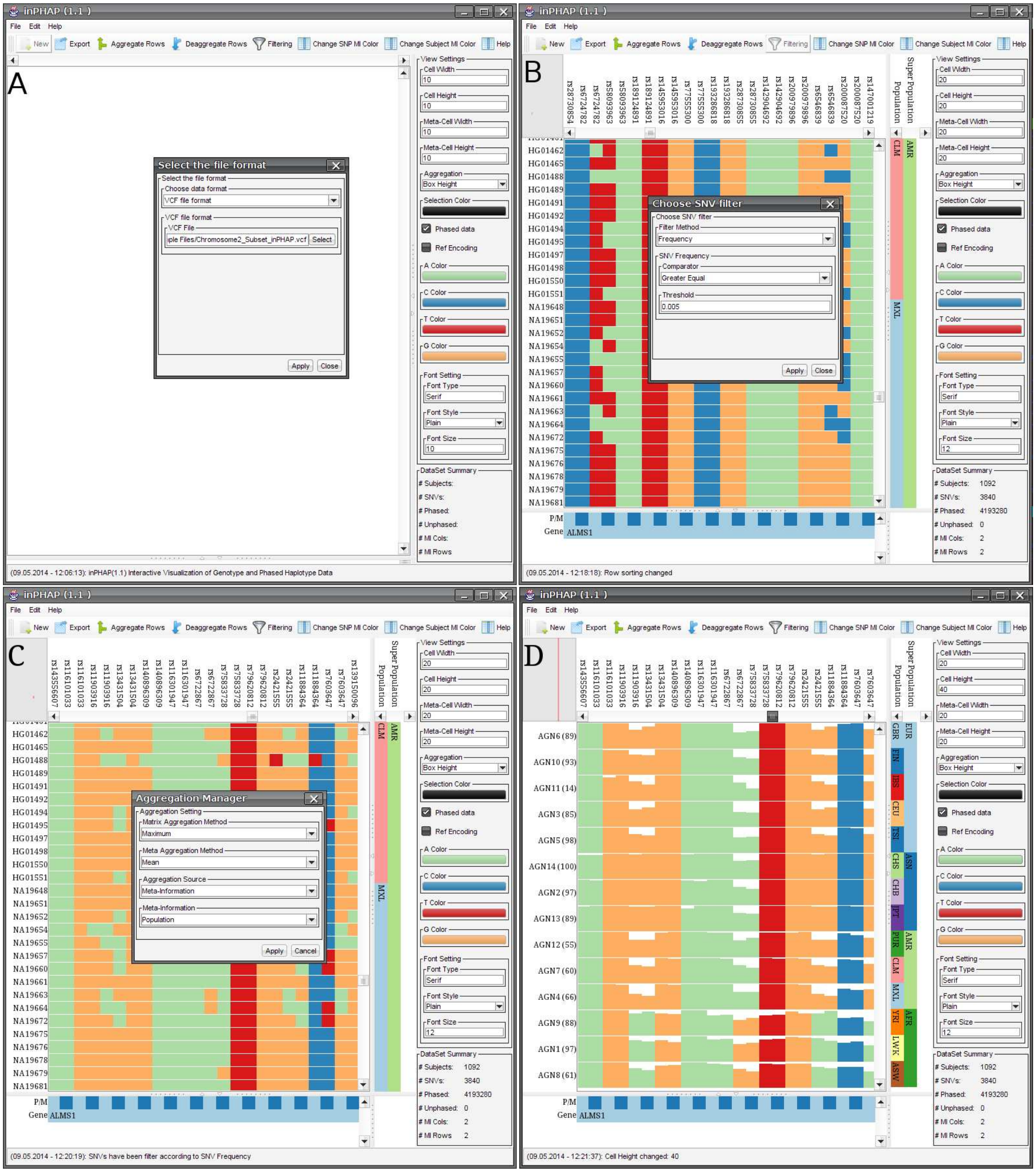

**Figure 3** Example workflow for the inPHAP tool, showing how data is loaded, processed and visualized using the inPHAP core features import, sorting, filtering and aggregation. **A**: The inPHAP graphical user interface after starting inPHAP and selecting "New" from the button menu on the top, in order to load a new data set in the VCF file format, **B**: View on the data, after loading a data set in the VCF file format and adding additional meta-information for individuals and SNVs in the data set. Rows have been sorted according to Population and Super Population by double-clicking the corresponding meta-information identifiers. "Filtering" from the button menu has been selected to initiate the filtering for SNVs with a frequency $\geq 0.5\%$, **C**: After filtering, the "Aggregate" button from the menu bar has been clicked to start aggregating the rows based on the provided meta-information. Here the population affiliation of the individual subjects is used for aggregation, **D**: Aggregated view on the filtered data set. In addition, zooming with the mouse wheel on the haplotype visualization was performed to increase cell height. The new height values are displayed in the settings panel.

The aggregation of haplotypes is accompanied by the aggregation of corresponding meta-information values. Meta-information can also be aggregated based on a user defined aggregation method which may differ from the method chosen for the haplotype visualization. In Figure 2 SNVs for the MLD associated gene *ARSA* are shown. Figure 2B shows the data after applying the minimum aggregation method to subjects that belong to the same population. This view is compared to a non-aggregated version showing the same data (see Figure 2A). After aggregation it becomes clearer, which SNVs are rare for specific populations, and how rare variants differ between the populations.

### Typical inPHAP workflow

An example workflow, showing how data is loaded into inPHAP, how filtering is applied to SNVs of interest and how aggregation is used to enhance visualization using meta-information is shown in Figure 3. This figure is split into four sub-figures showing the different stages of a typical inPHAP workflow. The quick button bar provides helpful features for processing the data. First data can be loaded into inPHAP with the “New” button. This opens up the settings dialog, from which the user can select what type of data he wants to load (see Figure 3A). As soon as data has been loaded (including meta-data), the user can interact with it, for example by sorting the rows based on meta-information. This can easily be done by double-clicking on one of the meta-information identifiers. To concentrate on SNVs of interest several different filters can be applied. Via the “Filtering” button in the quick button bar, the user gets access to the filter settings dialog, from which a filter of choice can be selected and parameters for the filter can be set (see Figure 3B). Data can be explored at any time, by navigating through the visualization using the corresponding navigation bars or by zooming in and out either with the mouse wheel or using the settings panel on the right of the graphical user interface. If needed, aggregation, e.g. based on meta-data, can be performed to obtain an aggregated view where individual subjects are grouped together based on the selected subject meta-information column and consensus values are calculated. This can be achieved by clicking the “Aggregate Rows” button from the quick button menu and setting up the corresponding aggregation parameters in the aggregation settings panel that shows up (see Figure 3C). The calculations for the aggregations are performed in the background, keeping the visualization usable at any time. A resulting view on the data after filtering, sorting, aggregation and zooming is shown in Figure 3D.

#### *Export*

With inPHAP the user can generate graphics in publication ready quality as either bitmapped images (PNG, JPEG and TIFF formats) or as scalable vector graphics (SVG or PDF format). During the export the user is provided with a preview of the resulting image as well as further options to adjust the image size. Furthermore, the user can decide whether to export the full visualization or just the region of the visualization currently visible in the inPHAP GUI.

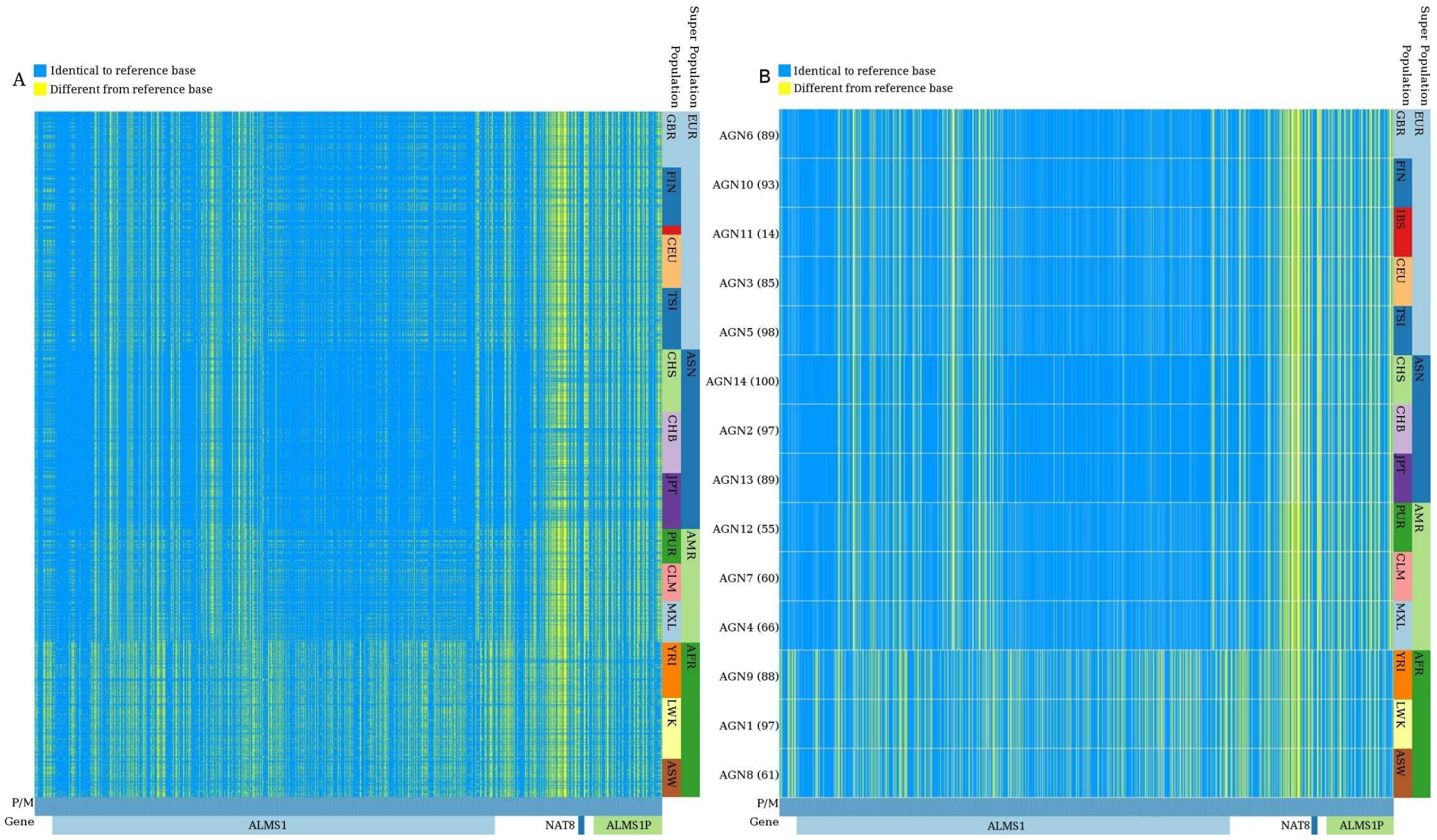

**Figure 4** inPHAP phased haplotype view for the 100-kb region on chromosome 2 spanning the genes *ALMS1*, *NAT8* and *ALMS1P*, after filtering of only those variants with a frequency $> 0.5\%$ across the 1096 human individuals of Phase 1 of the 1000 Genomes Project. SNVs that differ from the reference base are colored yellow, while alleles equal to the reference are colored blue. A: Individuals are sorted according to their affiliation with a common population. B: Shows the same visualization as in A, but individuals are aggregated based on their population affiliation, using “maximum” as aggregation method. For abbreviations of the population names see Figure 2.

## Results

### Visualization of genetic variation between populations

We applied inPHAP to haplotype data as generated by the 1000 Genomes Project. In the Phase 1 publication, Abecasis *et al.* provide a detailed view of the variation across several populations [24]. During their analysis they highlighted a 100-kB region on chromosome 2 spanning the genes *ALMS1* and *NAT8*. Variations in those genes have been associated with kidney disease in earlier studies [27]. As a proof of concept we used inPHAP to generate a similar visualization as Figure 2a in [24]. We first loaded the complete vcf file of chromosome 2 as provided on the ftp site of the 1000 Genomes project website. We then filtered only the respective 100 kB chromosomal region of the two genes. Next we applied two SNV filters: one for variants with a frequency $> 0.5\%$ across all individuals and one for rare variants with a frequency $< 0.5\%$. The resulting inPHAP visualizations are shown in Figure 4 for variants with a frequency $> 0.5\%$ and in Figure 5 for rare variants with a frequency $< 0.5\%$. As in Figure 2a of Abecasis *et al.* differences in common single nucleotide variants between different populations are clearly visible. Especially in the African (AFR) super population there are substantially more SNVs in the *ALMS1* region than for the other populations. This effect is even more pronounced after aggregation (see Figure 4B). Interestingly, for the Asian (ASN) population only very few variants are found in the central part of the *ALMS1* gene, while these are more likely in Europeans (EUR) and Americans (AMR). In contrast to all the other populations variant locations in this 100-kb region are more uniformly distributed, while for the other population groups variants are located mainly across two different sub-regions, namely the first part of the *ALMS1* gene and an approximate 20-kb region at the end of the selected 100-kb region spanning the genes *NAT8* and *ALMS1P*. These observations correlate well with the findings of Abecasis *et al.*, who showed that highly frequent variants in the 100-kb region are differently distributed across several populations.

Taking a closer look at the rare variants with a frequency $< 0.5\%$, one can see that the African population (AFR) again shows a higher number of variants than the rest (see bottom three rows in Figure 5). In addition, the degree of rare variants varies between dif-

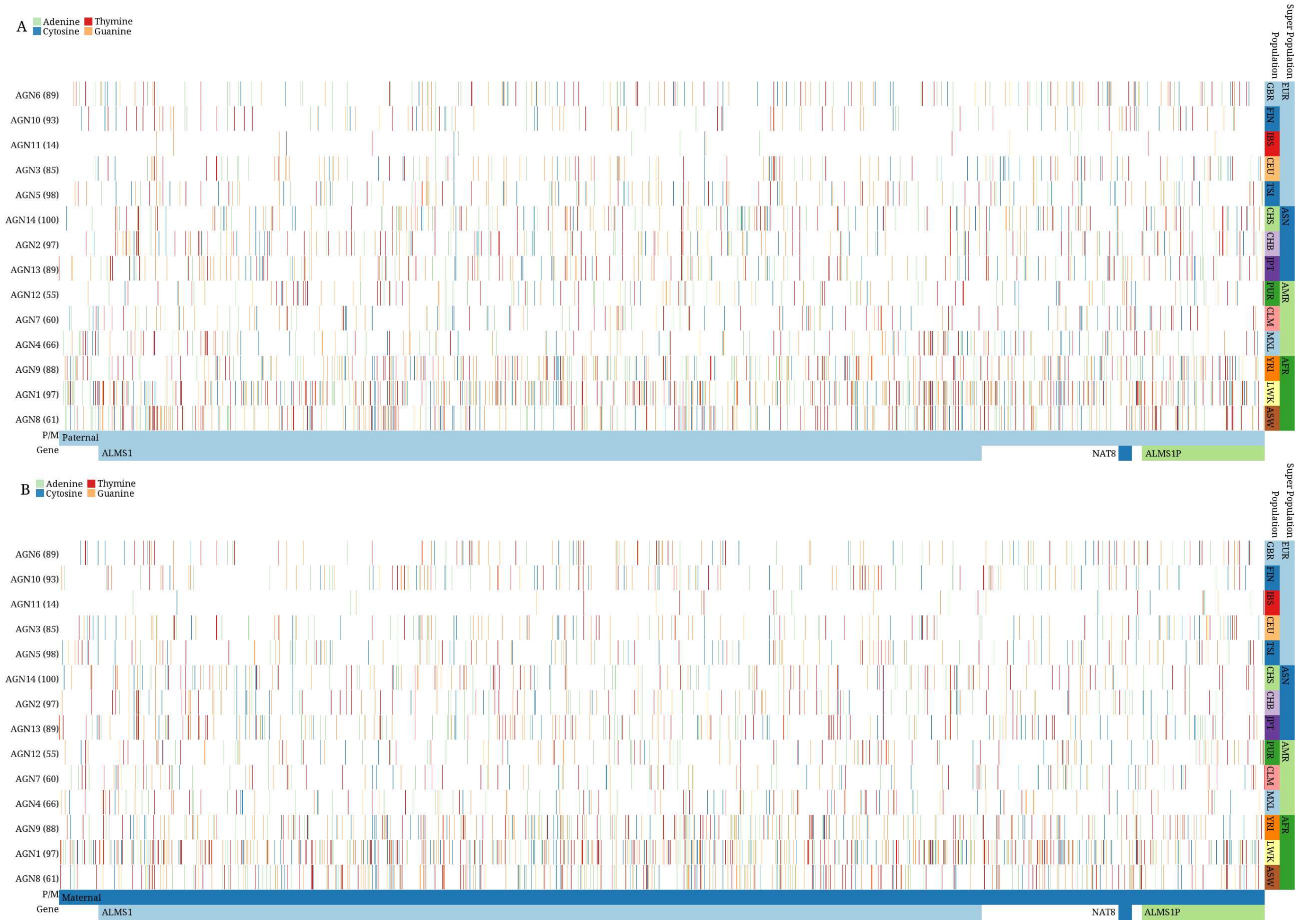

**Figure 5** inPHAP phased haplotype view for the 100-kb region on chromosome 2 spanning the genes *ALMS1*, *NAT8* and *ALMS1P*, after filtering of rare variants with a frequency $< 0.5\%$ across the 1096 human individuals of Phase 1 of the 1000 Genomes Project [24]. The bases A,C,T,G are colored green, blue, red and yellow respectively. Individuals are sorted according to their affiliation with a common population, and subsequently aggregated according to a specific population using the "minimum" aggregation method for SNVs. **A**: SNVs on the paternal chromosome are shown. **B**: SNVs on the maternal chromosome are shown. For abbreviations of the population names see Figure 2.

ferent populations, even for those from a common super population. For example, the Iberian population in Spain (IBS) shows only very few rare variants in this region (third row in Figure 5) whereas the numbers are much higher for the other European (EUR) populations. Interestingly, variations in the IBS population usually are limited to a single chromosome, which means that the SNV can either be found on the paternal or on the maternal chromosome, but rarely on both. This leads to the assumption that those variants have been introduced only recently, which correlates with the findings by Abecasis *et al.*, who argue that recent events, such as clan breeding structures or admixture of diverged populations are the main reason for rare variants in the Spanish (IBS) and Finnish (FIN) population [24].

## Visualization of MLD variations

Especially of interest for researchers are not common variants, that can be easily found in haplotype data, but rather rare alleles that can only be found in smaller subsets of populations or individuals. Finding such rare alleles can be difficult, due to the total number of subjects in common haplotype datasets, that might not include individuals with such rare alleles and furthermore the difficulty to filter out common alleles that are not as alluring as rare ones. inPHAP provides different methods in order to ease the search for rare alleles in large haplotype datasets, such as the frequency filtering feature together with the powerful aggregation methods included in the tool.

Metachromatic leukodystrophy (MLD) is an inherited disorder, that directly effects the growth and development of myelin, which is a crucial insulator around nerve fibers in human central and peripheral nervous systems [28]. The disease is caused by several missense mutations on Chromosome 22, causing defects of the enzyme arylsulfatase A ($ARSA$) [29]. One of the SNPs with dbSNP ID `rs743616` that is the supposedly responsible mutation for MLD, is a $C \rightarrow G$ substitution, leading to an amino acid change of Threonine $\rightarrow$ Serine in the corresponding protein $ARSA$. Using inPHAP we aggregated the dataset of chromosome 22 according to the population and then compared the resulting aggregations with respect to their super populations. Interestingly, as can be seen in Figure 6, there exist differences between super population groups, for example the Asian (ASN) and African (AFR) super populations show low pathogenic allele counts for MLD, whereas the European (EUR) and American (AMR) super populations show significantly higher total counts of pathogenic alleles, most pronounced is the Puerto Rican (PUR) population group in the American super population. On the single individual level the variations between subgroups are difficult to spot, as the allele patterns themselves in populations look entirely random without the aggregation. After aggregation in inPHAP the pattern becomes nicely visible. Furthermore, with this visualization the origin of the corresponding (pathogenic) allele can be distinguished, as both maternal and paternal allele frequencies can be observed in our visualization. For example for this SNP it seems to be of mostly maternal origin for Mexican individuals living in Los Angeles (MXL), as can be seen in Figure 6 (bottom fourth row).

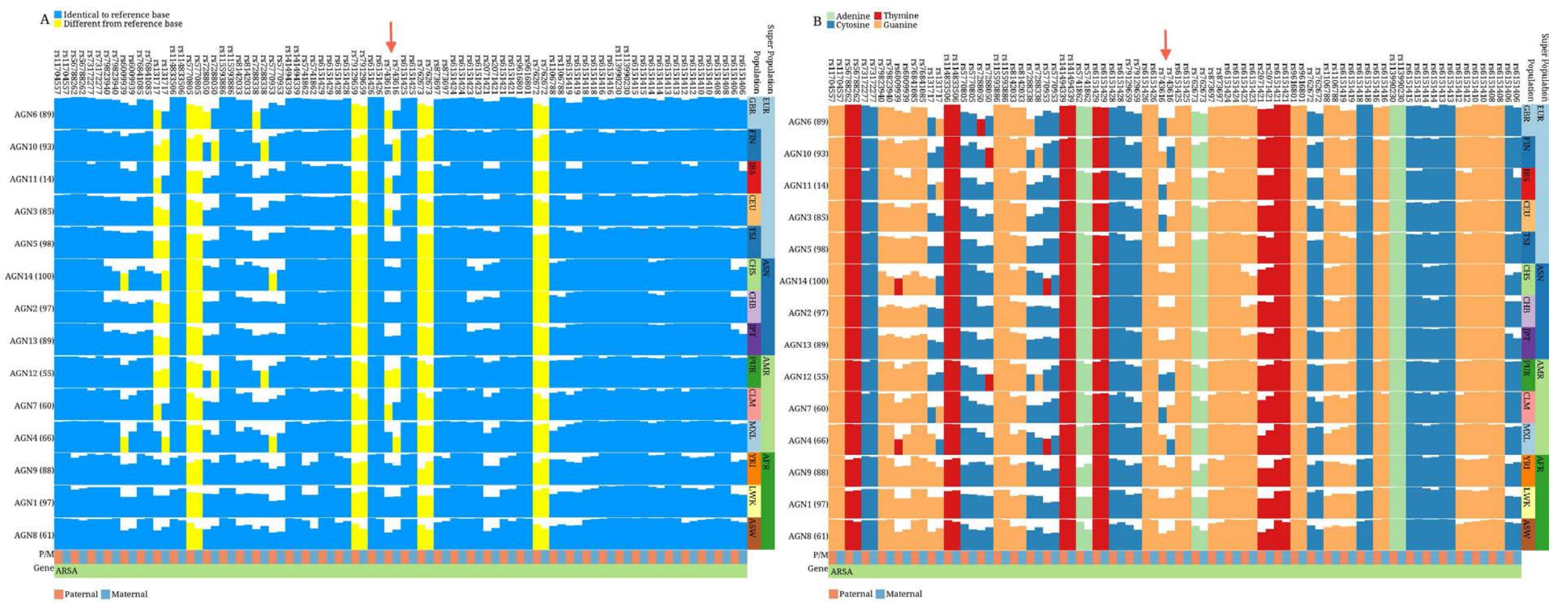

**Figure 6** Two inPHAP visualizations showing SNVs for the MLD associated gene *ARSA*. SNVs have been filtered based on their frequency across the 1096 individuals in the data set, showing only those SNVs with a frequency $> 0.5\%$. Individuals have been aggregated according to their population affiliation (for abbreviations of the population names see Figure 2). Bar heights for each SNV display the frequency of the aggregated consensus base. The arrow points to the maternal allele of the central SNV with dbSNP ID `rs743616` which is assumed to be one of the causative mutations leading to MLD. **A**: shows the selected SNVs using the reference- based visual encoding. Four of these SNVs show large differences to the reference base, which is shared across all populations, indicated by yellow bars ranging over a whole column, while some SNVs differing from the reference are restricted to few or even single populations, **B**: shows the selected SNVs using the nucleotide-based color encoding. In both visualizations differences between maternal and paternal alleles stand out clearly.

### Performance

The inPHAP tool has been designed in a way to keep the performance on a high level throughout the whole analysis. As an extreme use case, we tested inPHAP with the VCF file of chromosome 2 from Phase 1 of the 1,000 Genomes Project with 3.2 Mio SNVs and 103 GByte file size (for the VCF). inPHAP needs about 21 Gbytes of RAM, which can be explained by the fact that besides the raw allele data, all images are held in RAM as well. inPHAP still remains interactive and reacts smoothly when users switch between views, or apply functions such as filtering or aggregation.

## Discussion

We have designed inPHAP as a tool following Ben Fry's computational information design approach to understand large and complex data [30], which consists of the following seven main steps: acquire, parse, filter, mine, represent, refine and interact. With inPHAP, data can be loaded from different files formats, several filters can be applied, aggregations can be calculated, different representations for the underlying data are available, data can be sorted according to meta-information and interaction is possible at every stage of the analysis.

inPHAP can visualize phased haplotype data in order to study the influences of certain alleles. This is achieved by introducing two columns for SNVs, one for the maternal and one for the paternal allele. This design choice was motivated by the 1000 genomes data from Abecasis *et al.* [24]. Although inPHAP is designed for diploid organisms, its visualization concept can easily be extended to more complex genomes, as for example from omniploid organisms, by extending the number of columns used for single variations. Furthermore, the decision to split SNVs that are located on different homologous chromosomes into two different columns in the visualization has several further advantages. First of all, visual clarity is maintained throughout the whole analysis and comparisons between SNVs on homologous chromosomes can easily be made, by placing them next to each other in the haplotype visualization. Additionally, reordering of SNVs based on meta-information, such as the affiliation with a chromosome, enables the user to study single haplotypes without getting confused by the information from other homologous chromosomes. This would for example not be possible by adding two rows for each subject, as it was done in Figure 2a from Abecasis *et al.* [24], rather than adding two columns for each variant in the case of a diploid organism. In addition, comparison of haplotypes on homologous chromosomes is much easier, when the corresponding variations can be placed into chromosome based groups. The drawback of this approach is that comparison between patterns on the paternal and the maternal chromosome can become difficult, when the haplotype regions are large. In order to identify patterns on the paternal or maternal allele one would have to sort the SNVs according to their allele affiliation. However, this places maternal and paternal haplotypes far away from each other in the inPHAP visualization panel. Due to a limitation in the users screen size displaying both, the maternal and paternal haplotype, at the same time would be impossible. This could be overcome, by allowing the user to split the haplotype visualization panel in such cases into two parts, one for the maternal and one for the paternal allele, which is however currently not possible. Although, inPHAP was designed for phased data, it is not limited to those and can easily be applied to unphased data as well. Then of course, only a single column in the haplotype visualization panel is needed.

The possibility to decide whether specific allele combinations have an influence on an individual phenotype, is of great advantage and may lead to more precise interpretations. For this, we have shown that aggregations are a valuable tool to assess hidden patterns in the data and thereby help the user to draw better conclusions. However, aggregation techniques also bear risks. Depending on the aggregation method that is used, valuable information is potentially lost. During the analysis of the Abecasis *et al.* data set we have shown how aggregation can be used to display specific patterns hidden in whole populations. However, using a single aggregation technique did not allow us to reveal all the hidden information. Using the maximum aggregation technique, for example, enabled the comparison of common features, but has the disadvantage of loosing information on rare variants. In order to concentrate on rare variants, we had to apply the minimum aggregation technique. Therefore, the question whether to use aggregation for data exploration and which aggregation method is applicable, largely depends on the data and the question one wants to solve. For the visual encoding of aggregations we have implemented two different alternatives, a saturation based approach and the possibility to display nucleotide frequencies by using bars of different height. Using bar heights has the advantage that aggregated frequencies are much easier to compare between specific SNVs of interest. However, for a more general overview, e.g. over a whole genomic region, using saturation is more suitable, because depending on the number of SNVs and aggregated sub-groups in the overview, nucleotide boxes can become very small.

The application of different visualization strategies requires to be able to switch between data transformations and visual representations interactively. Since visualizing too much information in a single view easily leads to unnecessary clutter, which exacerbates the process of making decisions on the data, we follow a different strategy in inPHAP. By offering the user a variety of visual encodings and interaction techniques to process the data, he can generate different views on the data and switch between them in a fast and interactive way. In inPHAP we provide two different visual encodings for phased haplotype data, a reference based encoding where only similarities with and differences to the reference nucleotides are displayed and a nucleotide based representation that provides detailed base information. Only by the interplay of these two representations one is able to locate SNVs of interest and get nucleotide information at the same time. Again, in order to compare different representations, it would be of advantage to place them next to each other, which is currently only possible by exporting the visualized data using one of the available image formats in inPHAP. However, with that approach interactivity would be lost.

In addition to the visualization of phased haplotype or genotype data, meta-information, such as gene affiliation of SNVs or population information for individuals can provide further insight into the data.

So far inPHAP supports numerical and categorical meta-information for SNVs and individuals. Due to the generic design of meta-information for subjects, inPHAP can also handle quantitative meta-information, enabling the study of QTLs (quantitative trait loci) or eQTLs (expression quantitative trait loci). However, more complex meta-information, such as SNV associations, structural variations or individual relationships, can currently not be visualized without larger modifications of the tool itself.

### Future Work

An important step to evaluate and improve inPHAP will be the execution of a user study, which we will conduct next. Furthermore, we plan to improve inPHAP by adding more features. First of all we will add an additional component to the GUI showing the location of variations on the chromosome. This helps to identify SNVs in close proximity to each other which is of interest, since those variants are more likely to be in linkage disequilibrium. A further step in this direction would be to include additional visualizations in inPHAP, as for example an interactive LD-plot that can be linked to the haplotype visualization panel to improve identification and assessment of LD blocks. But also statistically motivated visualizations, such as charts that display the SNV frequencies for specific subgroups can largely improve inPHAP's efficiency, by making it easier to estimate differences between these groups.

In the current version of inPHAP we concentrated primarily on single nucleotide variations. However, also insertions and deletions (INDELs) are important variations that can lead to changes in gene function and consequently to disease. In future versions, we plan to extend inPHAP to be able to visualize INDELs together with SNVs, by adding a separate visual encoding for INDELs. Since INDELs can also differ between the maternal and paternal chromosome, the general concept of representing phased variations in different columns does also apply.

To improve interactivity with the visualization we also plan to add the possibility to keep user-defined regions in the visualization fixed, such that those regions are presented to the user at any time. In this, one would be able to navigate through the visualization in order to compare structures at different locations to the fixed region more easily. Another possibility would be to allow the user to split the haplotype visualization panel and link the resulting two sub-panels to each other, such that navigating in one panel would also change the view in the other panel. With this strategy interactivity would be maintained at any time.

## Conclusion

We have presented inPHAP, a tool for the visualization and interactive exploration of phased haplotype data for large scale genome projects. Through a variety of different interaction and data transformation possibilities, inPHAP allows the user to study the influences of variants either on the individual level or on a more general level that can for example be defined by meta-information. Since identical genotypes may have different impact, depending on their phase, visual assessment of the phase information can help researchers to make well-informed decisions. To our knowledge inPHAP so far is the only available interactive visualization tool capable of visualizing phased haplotype data.

#### List of abbreviations used

ARSA: *Arylsulfatase A*, COGS: *Collaborative Oncological Gene environment Study*, eQTL: *expression Quantitative Trait Locus*, GUI: *Graphical User Interface*, GWA: *Genome Wide Association*, GWAS: *Genome Wide Association Study*, HLA: *Human Leukocyte Antigen*, INDEL: *Insertion / Deletion*, MI: *Meta-Information*, MLD: *Metachromatic leukodystrophy*, SNP: *Single Nucleotide Polymorphism*, SNV: *Single Nucleotide Variation*, QTL: *Quantitative Trait Locus*

#### Competing interests

The authors declare that they have no competing interests.

#### Authors contributions

GJ and KN extended the general idea of iHAT to phased data and developed the concept of inPHAP. GJ designed the graphical user interface of inPHAP and the different visual encoding strategies. GJ and AP implemented inPHAP in the Java programming language. GJ, AP, and KN analysed the Abecasis *et al.* data set. AP, and KN investigated the genetic variations at the *ARSA* gene locus. All authors wrote, read and approved the final manuscript.

#### Acknowledgements

We acknowledge support by Deutsche Forschungsgemeinschaft and Open Access Publishing Fund of Tübingen University.